# MODELING OF ATMOSPHERIC FLOW AROUND A COASTAL CAPE: LEE SIDE STORY

Natalie Perlin[1,2], Eric D. Skyllingstad[2]

[1] – corresponding author, nperlin@coas.oregonstate.edu

[2] – College of Earth, Ocean and Atmospheric Sciences, Oregon State University, 104 CEOAS Admin Bldg, Corvallis, OR 97331-5503

# ABSTRACT

The current research focuses on mesoscale dynamics of the atmospheric circulation around an idealized coastal cape representing typical summertime circulation along the northwest coast of the U.S., studied using a mesoscale coupled ocean-atmosphere modeling system. The orographic wind maximum features a strong NW flow extending a few hundred kilometers downstream and seaward of the cape, which closely follows mesoscale orographic low pressure developed in the lee of the cape. Both wind maximum and the lee trough experience a pronounced diurnal cycle, marked by maximum northwest flow and minimum pressure in the local evening hours (its opposite phase during morning hours), and confirmed by observations from limited buoy and coastal stations.

Vertical structure of the atmospheric boundary layer over the coastal ocean on the lee side of the cape indicated the *downward* propagation of potential temperature and wind features during the course of the day, as opposed to the traditional surface-driven development of the atmospheric boundary layer. Momentum analysis showed the local pressure gradients near the surface vary greatly depending on the relative location about the cape. Strong perturbations in relative vorticity downwind of the cape supported the hypothesis that mountain dynamics have a key role in establishing wind regime around the cape. The sensitivity studies indicated the importance of coastal terrain elevation on the formation and the strength of the lee trough and wind maximum. Higher coastal topography was found to cause greater separation of the upwelling front (jet) from the coast downwind of the cape.

# 1. Introduction

The differences in wind regimes upwind and downwind of coastal capes and points along the U.S. west coast have been reported earlier in both observations and numerical simulations, and are viewed as atmospheric flow adjustment to a coastal cape or bend (Beardsley et al., 1987; Winant et al., 1988; Burk and Thompson, 1996; Burk et al., 1999; Dorman et al., 2000; Edwards et al., 2001, 2002; Perlin et al., 2004, 2007, 2011). Downwind wind maxima are often referred to as "expansion fans", alternating with "compression bulges" upstream. The term "expansion fan" was borrowed from hydraulic theory, indicative of a supercritical flow that results as the atmospheric flow rounds the cape and becomes confined in a thinner marine boundary layer on the lee side. A supercritical flow occurs when the flow speed is greater than the speed of internal gravity waves, and thus their ratio, a Froude number, is greater than one. The presence of such a flow could indeed be often diagnosed downstream of capes (Haack et al., 2001). However, the signature of the atmospheric flow variations extends above the marine boundary layer (Burk and Thompson, 1996), and was also reported to undergo diurnal changes on the lee side of the cape (Perlin et al., 2004, 2011). Our study supports and further explores the hypothesis of primary importance of mountain flow dynamics in regulating the atmospheric regime around the cape, which further can have strong effects on the coastal ocean circulation.

Formation of a pressure trough on the lee side of large-scale mountain ridges is well known and referred to as "lee troughing". Its formation is primary explained by Ertel potential vorticity conservation, based on the balance between the relative vorticity, Coriolis parameter (planetary vorticity) and height of the air column (Holton, 1992).

Formation of a mesoscale lee side trough, however, could include additional factors such as adiabatic warming of the descending air on the lee side, differential heating of the slopes, and latent heat release from cloud evaporation (Whiteman, 2000). Vorticity generation in lee trough and mountain wave events have been studied by McKendry et al. (1986), Smolarkiewicz and Rotunno (1989), Rotunno and Smolarkiewicz (1995), Batt et al. (2002); Epifanio and Rotunno (2005). In Pacific coastal region, in particular, coastally trapped disturbances (CTD) and Catalina Eddies were found to have lee troughing as an essential dynamic factor (Davis et al., 2000; Skamarock et al., 2002), and indicated the importance of the marine boundary layer depth and ambient synoptic conditions in the formation of the orographic lee side effects. Idealized study of the flow past the 3-D obstacles (Smolarkiewicz and Rotunno, 1989) yielded a pair of vertically oriented vortices and a flow reversal on the windward side of the obstacle; they argued that vertical component of vorticity developed due to the tilting of baroclinically-produced horizontal vorticity. Our study applies vorticity analysis to emphasize the importance of lee side dynamics in the atmospheric flow over the coastal ocean.

Here we apply an atmospheric numerical model to simulate mesoscale phenomenon associated with small-scale orographic features at the ocean-land boundary. The goals of the present modeling study are the following:

1) Identify the major features of flow dynamics around the coastal promontory;

2) Validate the modeling results using similarities from the observations (buoy, coastal stations, and satellite images);

3) Identify factors shaping the wind regime in the atmospheric boundary layer over the water and over the land on the lee side of the cape;

70  4) Determine the role of terrain elevation in formation of lee side features, and its
71      effects on coastal ocean circulation.

72   The paper is organized as follows. Information on the components of the
73 modeling system, details and references to the simulation design are presented in Section
74 2. Section 3 focuses on temporal averages and diurnal variability of the atmospheric
75 properties, as well as model validation. Section 4 presents momentum analysis and
76 discusses vertical structure of the boundary layer in the region of interest. Vorticity
77 analysis and the major terms of the vorticity equation are presented in Section 5. Section
78 6 discusses sensitivity studies investigating the topography role in the atmospheric and
79 ocean circulation around the cape. Summary in Section 7 concludes the manuscript.

80

## 81 2. Set of numerical experiments using a coupled ocean-
## 82    atmosphere modeling system

### 83 *2.1. System components, model domain and experiment setup*

84   The fully-coupled ocean and atmosphere modeling system used in the study was
85 developed from an existing mesoscale atmospheric model, ocean model, and a coupling
86 software package. The atmospheric model comes from the atmospheric component of the
87 Coupled Ocean/Atmosphere Mesoscale Prediction System (COAMPS$^{TM}$), developed by
88 the Marine Meteorology Division (MMD) of the Naval Research Laboratory (NRL). This
89 three-dimensional modeling system is based on the fully compressible form of the
90 nonhydrostatic equations (Hodur, 1997) and is widely used for both short-term
91 operational weather prediction in various regions around the world, and research purposes. The ocean component is based on the Regional Ocean Modeling System (ROMS), a free-surface terrain-following hydrostatic ocean model (Song and Haidvogel, 1994; Shchepetkin and McWilliams, 2005), widely used by the scientific community for a diverse range of applications. A coupling software package, Model Coupling Toolkit (MCT; Larson et al., 2005) is based on a number of routines similar to message passing interface (MPI), to facilitate exchange of data between the two numerical models. Technical aspects of the coupling approach implemented in the current modeling system are presented in Warner et al. (2008).

The model domain represents an eastern ocean boundary along the north-south oriented coastline and features a single cape. The cape in the middle of the domain protrudes about 90 km seaward and extends about 350 km in alongshore direction. Simulations were designed to study coastal atmospheric and ocean circulation around coastal capes in a coupled system during wind-driven upwelling, and in particular, off the Oregon – northern California coast. This dictated a choice for model topography, bathymetry, initial, and boundary conditions, to reflect the state of the atmosphere and ocean during the typical summertime upwelling conditions in this region. The model was initialized with horizontally-homogeneous profiles of the atmosphere and ocean, and forced with steady atmospheric geostrophic winds of 15 m/s near the surface, reducing to 5 m/s above 1500 m above sea level (ASL). The ocean was initialized at rest with prescribed profiles of temperature and salinity. North-south periodic conditions are assumed for both ocean and atmospheric model components. Coupled simulations were conducted for 14 days (336 hours).

The present study is an extension of previous work reported in Perlin et al. (2011), where more specifics on numerical model setup, initial and lateral boundary conditions,

116  and rationale for experimental domain can be found. In the current paper, a more in-depth
117  analysis of the atmospheric part of the circulation is presented.

118

## 3. Key features of the atmospheric circulation around the cape

### *3.1. Modeled time-average properties*

122     Average 10-m winds from 14-day model simulation show north-northwesterly
123  flow, featuring a pronounced wind maximum over the ocean on the lee side of the cape
124  (Fig.1a). Northwesterly winds over the land are nearly 2-3 times weaker. The strongest
125  cross-shore gradients of average wind speed are located on the lee side of the cape,
126  because of the local ocean wind maximum in excess of 15 m/s being located within ~150
127  km distance from the local wind minimum ~3 m/s over the coastal slopes at the southern
128  edge of the cape. The weakest cross-shore gradients of the wind speed are located on the
129  upwind side of the cape where the flow is generally uniform.

130     The predominantly alongshore flow over the ocean initiates coastal upwelling
131  resulting in decreased sea surface temperatures (SST) along the entire coastline (Fig.1b).
132  In particular, lower SSTs and wider upwelling area are noted on the sea side of the cape.
133  The offshore increase of SSTs over the course of simulation are due to solar heating, as
134  well as the imposed incremental warming of the atmospheric profile in the model to
135  compensate for the typical large-scale subsidence during the summer. This heating
136  combined with the coastal upwelling produces a cross-shore SST difference of up to $6^{o}C$
137  on the downwind side of the cape by the end of the simulation. The location of colder water and the upwelling front help to explain increased atmospheric stability and decreased momentum transfer over these regions, discussed further in Section 4 dealing with momentum analysis and vertical structure of the lower troposphere.

Temporal average of the sea level pressure (Fig. 1c) indicates a pressure trough forming on the downwind side of the cape, as well as a weaker pressure ridge on the upwind side. Linear dimensions of the pressure trough on the lee side are similar to those of the entire cape. However, the trough is most pronounced over the coastal ocean, and is nearly non-existent over the land except along the southern slope. In contrast, the weaker pressure ridge appears mostly over the land, and its cross-shore inland extent is of a similar size to the alongshore extent.

Temporal average of height-integrated cloud water mixing ratio (Fig. 1d) gives an indication of cloud formation patterns around the cape. Strong preference in cloud formation is found on the upslope locations on the north side of the cape. Cloud-free regions are found on lee side of the cape collocated with the wind maximum. The latter effect is often observed in satellite images, as will be demonstrated later.

The prominent nearshore features on the lee side of the cape are apparent not only in temporal averages, but also in diurnal variability. Fourteen-day composites of surface winds and sea level pressure for distinct times of the day indicate that the near shore lee side feature undergoes a definite diurnal cycle (Fig. 2). The following convention is used for generality: 0600 LST is referred to as "morning" time, 1200 LST as "daytime", 1800 LST as "evening", and 0000 LST as local "night" time. Model northwesterly winds are up to 3-4 m/s stronger (weaker) than daily average in the evening (morning) hours on the lee side, closely corresponding to the area of coastal ocean where highest wind speed and sea level pressure anomalies are found, and in accordance of the deepening (relaxation) of the lee trough. Diurnal alongshore progression of the lee trough is evident from nighttime (daytime) locations of the pressure anomalies, in which the pressure anomalies move south of the cape, accompanied by ~ 2m/s strengthening (weakening) of northerly winds. For indication of diurnal progression of lee side pressure trough and wind maximum also see Fig. 7a and Fig. 9 in Perlin et al. (2011).

Despite the rather complex surface wind regime over the coastal ocean, winds over the coastal land surface within 50-100 km off the coast mostly follow the traditional sea-land breeze circulation. Thus, the strongest winds are found during the peak of solar heating hours, and the landward (eastward) sea breeze propagation is apparent from the positive wind speed anomalies seen in the early evening. The only exception to this general view is the narrow strip (less than 50km) on the SW slope of the cape, where complex interaction between the lee trough, shoreline orientation, daytime surface warming, and upslope-downslope (anabatic-katabatic) flows make the wind regime far more irregular and unique for each relative location on that slope.

Empirical orthogonal function (EOF) analysis of the surface perturbation Exner function in Fig. 3 reveals the modes of surface pressure variability. For the analyzed domain shown in the figure (slightly smaller than the modeled domain), EOF1 explains about 88% of the total variance, and EOF2 explains about 5% of the total variance, but the local contributions of each EOF vary around the cape. On the lee side of the cape, local contributions are around 80% for the EOF1, and up to 15-20% for the EOF2. Both EOFs indicate distinct pressure feature on the lee side of the cape varying diurnally, and correspond to deepening of the existing time-average pressure trough when the amplitudes are positive. Note that there is no zero crossing in EOF1, i.e., pressure change occurs over the entire domain simultaneously, but to a greater degree on the lee side of the cape. The zero crossing in EOF2 indicates opposite changes about the cape: pressure increases upwind of the cape over the coastal and coastal slopes while pressure decreases on the downwind side; the downwind pressure feature extends further inland. EOF2 may also partially reflect the effect of "topographical blockage" that is often considered responsible for a dipole feature on the upwind and downwind sided of a topographical feature (and is evident on the mean field in Fig.1c), but its variance contribution is smaller that EOF1, both in percentage and amplitude.

Being statistically orthogonal, EOF1 and EOF2 may nevertheless both outline similar processes that spatially evolve over the course of the day. This may explain peaking in one EOF amplitude time series occurring at the time of flattening of the other EOF amplitude, as between hours 140 and 164, or 164 and188. Downwind spatial extent of the lower pressure feature varies from about 280 km (EOF1) to about 350 km (EOF2), which is slightly smaller than the downwind extend of the average pressure trough (Fig. 1c).

Evidence of a significant diurnal cycle over coastal Oregon land surface was also found in Bielli et al. (2002). In the present study, we include the coastal ocean with specific focus on lee side locations. Atmospheric flow around the cape varies greatly not only spatially but temporarily, as shown by the diurnal cycle of wind speed and potential temperature in the lowest 1200m over selected locations along the coastline (Fig.4). A first general observation is that most of the locations experience a developing nighttime low-level jet (LLJ), marked by elevated wind maxima in the lower troposphere. This wind maxima is especially defined upstream of the cape and by the tip, and is more pronounced over the land than over coastal waters (L1, L2, O1, O2). Ocean point O2 at the traverse of the cape experiences somewhat persistent strong 17-21 m/s winds between 200-1200m. Diurnal temperature and PBL height (about 400m) variations are less prominent at the ocean points O1 and O2, with minimal or no diurnal heating of the air column. Diurnal land warming at L1 and L2 locations leads to PBL height increase to about 600m, which gradually occurs from the ground surface upward.

Diurnal evolution of the atmospheric thermal and wind structure on the lee side of the cape is notably more complex. Ocean location O3 shows diurnal warming of the air column propagating from the higher elevations *downward*. This warming starts in the morning and peaks around 1700 LST, at the time of maximum horizontal temperature gradient. Diurnal warming is accompanied by a strong elevated wind maximum at the top the shallower boundary layer of about 200m (see Fig. 4b in Perlin et al., 2011, for details on boundary layer analysis), reaching wind speeds of over 25 m/s at 1700-1800 LST. Relaxation of the vertical thermal structure after 1700 LST until midnight is also simultaneous with the vertical expansion of the wind maxima at O3 location. The thermal regime at O4 is marked by cooling of the air column at night and morning hours, and subsequent warming during the daytime; note that diurnal temperature variations are greater at 600-1200m than at lower elevations. The diurnal range of PBL layer height variations is up 200m, being greater than at other ocean locations. Low-level early evening wind maxima are not generated at the O4 location, where the near-surface winds are shown to be the weakest of all ocean locations.

229    Coastal land points (L3 and L4) on the lee side of the cape show weaker winds
230 and a less vertically organized thermal structure that extends to greater elevations,
231 evidenced by the increased diurnal PBL heights. Diurnal warming and cooling of the air
232 column does not necessarily occur from the surface upward, as in a simple case of
233 daytime boundary layer development over the land. The distinction of thermal and wind
234 regime on the lee side of the coastal ocean and land, marked by their variability
235 originating at higher elevations away from the surface, serves as further evidence of the
236 orographic influence on the flow circulation around the cape, and viewing the problem as
237 similar to a traditional mountain internal wave system.

238 *3.2. Model validation*

239    In order to validate these semi-idealized model simulations, we used observations
240 from the Oregon and northern California coast resembling the idealized model domain.
241 Numerous studies, and in particular, analyzing QuikSCAT satellite observations, have
242 already confirmed the existence of wind intensification features downstream of capes and
243 points as a permanent hallmark of the region. Diurnal variability has also been detected in
244 QuikSCAT observations (Perlin et al., 2004), but without a clear physical interpretation.

245    Multi-year observations of winds and atmospheric pressure from nearshore buoys
246 operated and supported by the National Data Buoy Center (NDBC) and NOAA's
247 National Ocean Service, provide a dataset for evaluating the key findings of the model
248 simulation. Several buoys were chosen for the analysis along the Oregon and northern
249 California coast along with the three coastal stations (Fig. 5), in the vicinity of the two
250 notable coastal promontories of Cape Blanco and Cape Mendocino. Each buoy or coastal

station represents a specific area relative to the nearest cape as follows. NDBC buoy 46015 represents the nearshore location at the traverse of the tip of the cape (Cape Blanco), buoy 46027 represents an area downwind of the cape (Cape Blanco), and the buoy 46022 represents the nearshore area upstream the tip of cape (Cape Mendocino). Coastal stations CARO3, PORO3, and CECC1 represent the coastal locations upwind of the cape, at the traverse, and downwind of the tip of the cape, respectively, all around Cape Blanco. A summary on the buoy and station information and the data used is shown in Table 1. Because the simulations were designed to represent summertime upwelling-favorable conditions, certain limitations were imposed on the observational data to conform to the model setup. First, only data from the upwelling season encompassing months May through September were used. Second, only days with no missing records were chosen. And third, the winds had to have a northerly component (v-component $< 0$) at all times to be included in the calculations. The length of the historical data record varied for each observational location, as well as the frequency of recorded observations (hourly, 10-min averages, or 6-min averages). Linear interpolation was used to fill small gaps (5 or less records) in the observational data. For each location except CECC1, the number of days analyzed was fairly long (greater than 400 - 800; see Table 1) providing a robust, statistically significant analysis. Note that coastal location of CECC1 station is such that expected development of diurnal sea breeze and upslope circulation would result in southerly component of the wind, i.e. v-component $> 0$. This eliminated a large portion of multi-year observations at this station from the current analysis; only 33 days passed the test of restriction to northerly wind component at all times.

The average diurnal cycle (Fig. 6) of 10-m winds is most evident for the downstream buoy (46027), where the generally strong winds speed increased over 50% in the early afternoon and evening compared to early morning times, from about 7 m/s to nearly 12 m/s. The buoy at the tip of the cape (46015) shows less than a 10% increase in wind speed at local noon as relative to the midnight value of 8 m/s. The upstream offshore location (buoy 46022) indicates weaker winds and somewhat reversed wind cycle: wind peaking in early morning, and being about 30% increase of the evening value of about 5.5 m/s. All the coastal stations indicate top wind speeds during mid-day, and weakest winds during the night or early morning hours. The downwind coastal station CECC1, however, indicates a wind increase of nearly 2.5 times from the morning to the late afternoon and evening values. Note that while sea breeze and coastal upslope winds are expected to contribute to the diurnal cycle at all of the coastal stations, the CECC1 location downstream of the cape demonstrates greater range of diurnal wind amplitude, and increased wind speed during later times than the other two stations upstream.

Observations of the atmospheric surface pressure at the same buoys and coastal stations were used to derive the pressure anomalies for these locations. The anomaly was computed as a departure from the 24-h running average sea level pressures. To ensure the similarity in observed weather conditions with the modeled conditions, surface pressure time series included in the analysis were subject to the same data constraints as the wind observations. All the stations showed a bimodal diurnal cycle in surface pressure anomalies, but with distinct variations characteristic for each location (Fig. 7). The buoy at the traverse of the cape (46015) shows two comparable negative anomaly pressure minima at 0400 LST and 1800 LST, and two positive anomaly pressure maxima at 1000-1100 LST and 2200 LST (the former slightly exceeding the latter). The downstream buoy 46027 shows a single negative anomaly pressure minimum at 1800 LST, which is more than twice the minimum at buoy 46015 at the same time. A single pressure maximum at the buoy 46027 occurs at 0800-0900 PST, being the strongest maximum among all the buoy records. The upstream buoy 46022 has a single pressure minimum at 0300 LST, and a positive anomaly, double pressure maxima around noon and 2100 LST.

Pressure anomaly amplitudes vary less among the coastal stations than among the buoy locations, but similar trends are nevertheless apparent: strongest amplitudes are at downstream location CECC1 with an early evening pressure minimum, and early morning negative anomaly magnitude is greatest at upstream location CARO3. The negative anomaly pressure minimum at 1800-1900 LST, however, results for all the coastal stations, unlike for the buoy locations.

These observations confirm the conclusions from the model results showing that the strongest winds and diurnal variability are found over the ocean on lee side of the cape, weaker winds and variability are found on the upwind side, and less diurnal variability is found in offshore locations near the tip of the cape. The wind observations confirm our findings about the modeled phases of the diurnal cycle in different locations relative to the cape over the coastal ocean, as well as indicating a distinctively traditional sea-land breeze circulation over the coastal land. Surface pressure observations support the presence of diurnal cycle, in which pressure minimums occur at opposite phases of the cycle in the lee side and the windward side, respectively. Additionally, the bimodal distributions of pressure anomalies suggest there might be more than one mechanism involved in shaping the diurnal variability.

The effect of cloud clearance downwind of capes and cloud build-up on the upwind slopes has been reported in earlier studies (Haack et al., 2001), and is clearly demonstrated in satellite images over the region of U.S. Pacific coast (Fig. 8). Both images correspond to the summertime conditions with corresponding northerly winds. Note that cloud-free areas are formed downstream of every prominent cape along the coast, especially as shown in the second image. The first image demonstrates enhanced cloudiness (denser and brighter white colors) both over the ocean and the coastal land upstream of the capes. Thus, modeled cloud formation and predominance around the cape (Fig. 1d) is very consistent with the observations in the same area.

## 4. Momentum analysis

Momentum budget analysis is a useful tool to study the dynamic equilibrium of the horizontal flow in the boundary layer and to determine the major contributing terms in the circulation. The vertical structure of the marine and coastal boundary layers over various locations along the U.S. Pacific coast had been estimated using moored observations and numerical models, in which spatial and diurnal variability in coastal flow was attributed to different balances of geostrophic acceleration, stress divergence, and pressure gradient terms estimated for cross-shore or alongshore momentum (Zemba and Friehe, 1987; Winant et al., 1988; Samelson and Lentz, 1994; Burk et al., 1999; Bielli et al., 2002).

Our study further investigates the dynamics of the coastal atmosphere by estimating the momentum balance terms in vertical profiles and at certain spatial locations that were found representative for the complex flow regime around the cape. Momentum components were computed in the model at every time step, and averaged hourly for the output. Major components of the momentum balance in u- and v-directions are the following:

$$\frac{\partial u}{\partial t} = -u\frac{\partial u}{\partial x} - v\frac{\partial u}{\partial y} - w\frac{\partial u}{\partial z} - \frac{1}{\rho}\frac{\partial P}{\partial x} + vf - \frac{\partial \overline{(u'w')}}{\partial z} \quad (1)$$

$$\frac{\partial v}{\partial t} = -u\frac{\partial v}{\partial x} - v\frac{\partial v}{\partial y} - w\frac{\partial v}{\partial z} - \frac{1}{\rho}\frac{\partial P}{\partial y} - uf - \frac{\partial \overline{(v'w')}}{\partial z} \quad (2)$$

I  II  III  IV  V  VI

where the numbered terms are the following:

I – storage terms;  II - horizontal advection; III – vertical advection; IV – horizontal pressure gradient; V – Coriolis term; VI –vertical divergence of turbulent momentum flux, i.e., vertical mixing term. Numerical diffusion terms are small and omitted in the analysis. Storage terms are reduced to zero when averaged over the longer time period. Horizontal pressure gradient is estimated from the two modeled components, one is imposed by the model as horizontally homogeneous and time-invariant and determined from the prescribed geostrophic wind, and the other is estimated as a remaining term after all other terms had been determined. The latter is further referred to as a "local pressure gradient", a departure from the geostrophic or synoptic-scale forcing. Thus, the modeled pressure gradient terms are $\frac{1}{\rho}\frac{\partial P}{\partial x} = v_g f + \frac{1}{\rho}\frac{\partial P_{loc}}{\partial x}$ and $\frac{1}{\rho}\frac{\partial P}{\partial y} = -u_g f + \frac{1}{\rho}\frac{\partial P_{loc}}{\partial y}$, in u- and v-directions, respectively, where $P_{loc}$ is a departure from a total pressure from the geostrophically balanced value. Using these definitions, the momentum balance equations for the modeled atmospheric flow become as follows:

$$-u\frac{\partial u}{\partial x} - v\frac{\partial u}{\partial y} - w\frac{\partial u}{\partial z} - \frac{1}{\rho}\frac{\partial P_{loc}}{\partial x} - v_g f + vf - \frac{\partial \overline{(u'w')}}{\partial z} = 0 \quad (3)$$

$$-u\frac{\partial v}{\partial x} - v\frac{\partial v}{\partial y} - w\frac{\partial v}{\partial z} - \frac{1}{\rho}\frac{\partial P_{loc}}{\partial y} + u_g f - uf - \frac{\partial \overline{(v'w')}}{\partial z} = 0 \quad (4)$$

in u- and v-directions, respectively.

Vertical profiles of 10-day average momentum balance terms at locations O2-O4 and L2-L4 (marked in Fig. 1c) around the cape are shown in Fig. 9. Upstream points O1 and L1 are of a less interest and are not presented. As a general remark, positive values of u-momentum balance terms contribute to increased westerly wind component, while at the same time, positive values of v-momentum balance contribute to *decreased* northerly (negative **v** values) wind component. Imposed initial and boundary condition dictate the geostrophic pressure gradient in u-direction, corresponding to 15 m/s winds below about 1500m, tapering to 5 m/s above that elevation, and zero pressure gradients in v-direction. Notice that due to terrain elevation of 750 m, all land points show the gradual decrease of pressure gradient for u-momentum at lower elevations above the surface than that for the ocean points.

The location O2 at the tip of the cape shows simple balance in u-momentum terms primarily between geostrophic pressure gradient, vertically uniform local pressure gradient, Coriolis term, and vertical mixing near the surface below 200m. Horizontal advection acting to decelerate the flow is weak. The momentum balance for the v-component at the same location indicates the primary balance is between a stronger local pressure gradient that strengthens northerly flow and two counteracting terms: horizontal advection and vertical mixing within the boundary layer. At elevations above 1400m, vertical advection acting to decelerate the flow also comes into play. The Coriolis term is weakly negative within the boundary layer, and weakly positive above it.

Ocean location O3 in the lee of the cape is close to the location of wind maxima, and it displays a local pressure gradient term in the u-direction that exceeds the geostrophic forcing within the boundary layer. This local forcing term gradually reduces with height to a negligible value at about 1300m elevation. All other terms mostly act to decelerate the westerly flow, with the turbulent vertical mixing term being the strongest near the surface. In the v-direction, vertical mixing is the primarily counterbalancing term within the 200-m boundary layer, while the local pressure gradient, Coriolis term and vertical advection are strengthening the northerly flow.

Further downstream, at the O4 location, local pressure gradient terms in both directions change sign to oppose the flow direction. Horizontal and vertical advection terms become important, often balancing each other. Their sum, however, remains largely positive below 600m (negative below 1000m) for the u-component (v-component), strengthening the flow. The notable feature of the O4 location is a greatly reduced turbulent vertical mixing term, as compared to other upstream locations. This apparently results from the upwelling-driven cold SSTs found in this region, and in the absence of the strong vertical shear (Compare locations of O3 and O4 in Fig. 1c with SST in Fig. 1b, and vertical wind profiles in Fig. 4 for O3 and O4 panels).

The land points L2, L3, and L4 all have a local pressure gradient that is acting to decelerate (accelerate) the flow in the lowest 1000-1500m in the u-direction (v-direction), except at the shallow layer near the surface at L4 for u-momentum balance and a weak pressure gradient term in v-momentum balance. In the absence of cold lower boundary conditions, vertical mixing is reducing the flow within the boundary layer for u-direction terms, and up to about 1000 m height for the v-direction terms. Horizontal and vertical advection terms are prominent at the L2 and L3 locations, except weaker vertical advection in u-direction for L3. These terms are notably smaller for the L4 location, where the local pressure gradient terms are also the weakest of all the land locations.

Diurnal evolution of the local pressure gradient terms at the surface (Fig. 10) shows, at times, opposite behavior of the u-components for the ocean and land points:

diurnal minimum for ocean points in afternoon/early evening correspond with diurnal maximum for the land points (O2 and O4 vs. L2 and L4). Locations on the lee side of the cape, O3 and L3, differ in another way: a strong positive peak results at the O3 location in the afternoon, which is three times the corresponding morning value, but the land location L3 lacks the definite afternoon extreme, showing about 60% weaker negative values as compared to the nighttime values. For the v-component of the surface pressure gradient term, both land and ocean locations indicate a similar diurnal evolution, especially in showing the extreme values at 1800-2000 LST that correspond to northerly winds increase.

The vertical analysis of momentum balance terms helps explain the diurnal behavior of the coastal cape flow field. First, our analysis shows the presence of strong local pressure gradients, acting to accelerate or decelerate the flow depending on relative location and proximity to the tip of the cape. These local pressure gradient terms vary diurnally, and peak in afternoon and evening hours because of surface land heating Secondly, strong horizontal and vertical advection terms often counteract each other and are highly dependent on a location around the cape. And third, notably reduced turbulent mixing is evident over the cold SST on the lee side of the cape and downstream of the major wind maximum.

## 5. Relative vorticity analysis

Vorticity analysis in our study aimed to emphasize the importance of mountain flow regime resulting in the lee side of the simulated topographic obstacle over the coastal ocean. Relative vorticity has three components in 3D space, yet its vertical 448  component, $\zeta = \frac{\partial v}{\partial x} - \frac{\partial u}{\partial y}$, is most interesting for air-sea interaction problems. Hereafter,

449  we will refer to the term "vorticity" as the vertical component. In the presence of surface

450  friction, the vorticity equation can be obtained by taking a derivative $\frac{\partial}{\partial y}$ of the Eq.(1), a

451  derivative $\frac{\partial}{\partial x}$ of the Eq.(2), and then subtracting the former from the latter. After

452  regrouping the terms and substituting $\zeta$ when possible, using full derivative to combine

453  storage and advection terms as $\frac{D\zeta}{Dt} = \frac{\partial \zeta}{\partial t} + v_i \frac{\partial \zeta}{\partial x_i}$, and assuming $\frac{Df}{Dt} = v\frac{\partial f}{\partial y}$, we yield

454  the following:

455  $\frac{D}{Dt}(\zeta + f) = -(\zeta + f)\left(\frac{\partial u}{\partial x} + \frac{\partial v}{\partial y}\right) - \left(\frac{\partial w}{\partial x}\frac{\partial v}{\partial z} - \frac{\partial w}{\partial y}\frac{\partial u}{\partial z}\right) + \frac{1}{\rho^2}\left(\frac{\partial \rho}{\partial x}\frac{\partial P}{\partial y} - \frac{\partial \rho}{\partial y}\frac{\partial P}{\partial x}\right) +$

456  $+ \left(\frac{\partial Fr_y}{\partial x} - \frac{\partial Fr_x}{\partial y}\right)$ , (5)

457  where the $Fr_x = \frac{\partial(\overline{u'w'})}{\partial z}$, $Fr_y = \frac{\partial(\overline{v'w'})}{\partial z}$; $v_i$ denotes velocity components $u, v, w$ in

458  corresponding $x_i$ directions $x, y, z$.

459   The Eq.(5) describes the terms resulting in changes in absolute vorticity, $(\zeta + f)$

460  on the left-hand side, as follows. The first term on the right-hand side (RHS) is vorticity

461  divergence term, often called "vortex stretching" term. The second term on the RHS is

462  tilting/twisting term; the third term is a solenoidal term, and the last one is turbulent flux

463  divergence term.

464   Temporal averages of vorticity and major terms of the RHS of Eq.(5) were

465  computed for two cross-sections from hourly model output, and then averaged over the

466  entire period of the simulation (Fig. 11). Modeled variables on terrain-following vertical coordinate system were interpolated onto level surfaces prior to estimating the relative vorticity and absolute vorticity equation terms. The vorticity estimates are smoothed using 5-point averaging as follows:

$$\hat{V}(i,j) = 0.2 * [V(i,j) + V(i+1,j) + V(i-1,j) + V(i,j+1) + V(i,j-1)],$$ where $\hat{V}$ is the smoothed variable estimated using values of $V$ at five locations given by horizontal indices $i, j$.

Alongshore cross-section (Fig. 11, top left) shows elevated relative vorticity near the surface within a few hundred meters on the north of the cape, and notably greater vertical propagation of positive vorticity on the south (downwind) side. The elevated maximum on the lee side is separated from the water surface, being an extension of a lee-side coastal maximum. Negative values are found at higher elevations on the upwind side, but not on the lee side. Higher positive vorticity in the cross-shore (top right panel) direction is found over the coastal slope, separating from the surface over the coastal water and extending vertically about 50 km off the coast. Stronger negative vorticity appears above the topographic inflection point. Vorticity gradually decreases away from the topographic change, with the downstream lee side extent on the order of hundreds of kilometers.

The temporal average of the relative vorticity term shows the mean spatial picture of the vertical distribution. Temporal variations are caused primary by the divergence term, tilting/twisting term, and turbulent flux divergence term. The temporal mean of the baroclinic term is about 100 times smaller than the others, and is not shown. All of the three terms shown (Fig 11, second-forth row) show very strong spatial variability over the topographic changes, and extent further on the lee side of the cape. Out of three terms, the vorticity divergence appears to contribute the most in formation of spatial structure of the relative vorticity in downstream direction (alongshore cross-section) and elevated maxima within 50 km off the coast (cross-shore cross-section). This basic vorticity analysis confirms the strong influence of the topography on the flow extending hundreds kilometers downstream on the lee side of the cape.

## 6. Sensitivity studies

We conducted sensitivity studies with three additional experiments to examine the influence of the elevation of the orographic coastal barrier on the atmospheric circulation. The experiments were similar to the control case in all but the maximum topography elevation. In control case, topography increased at a rate of 25 m of elevation per kilometer of inshore distance, and remained flat eastward after it reached 750 m. In the first sensitivity study, we set flat topography at 0 m. In second case, topography increased at the same rate as in the control case, but only up to 375 m, a half of that of the control case. In the third case, the maximum elevation was set to 1500 m, twice of the control case maximum.

Average surface winds and sea level pressure for the three cases, as well as the resulting SST changes are shown in Fig. 12. In no-topography case, only barely noticeable pressure ridge formed over the upstream side of the cape, very little weakening of the flow over the coastal ocean; no apparent wind maximum was formed on the lee side. North of the cape and sufficiently far downstream of it, surface winds within ~50km offshore are the strongest for this case out of all simulations, due to the absence of any topographic blockage. Strongest coastal winds also lead to strongest nearshore upwelling out of all simulations along the straight coastline and southern part of the cape, exceeding 4°C of SST drop.

Simulation with topography maximum at 375 m shows weak lee trough and the beginning of a wind maximum. Weaker nearshore winds along the rest of the coastline diminish coastal upwelling compared to the flat-land simulation, except on the downwind side of the cape. The upwelling front outlined by negative SST differences, starts to separate off the coast south of the cape in this simulation, yet its predominant structure is following the coastline. Note that the sea level pressure and surface winds fields along the lateral boundaries are very similar in the above two cases (Fig. 12 a-b), and only vary slightly in the control case (Fig. 1 a-c).

A very different scenario results for the third case (Fig. 11b) with the 1500-m coastal barrier, which forces notable variations across the entire domain including the lateral boundaries. A deep lee side trough is produced south-southeast of the tip of the cape, which forms a separate low pressure system with a strong wind. The lee side feature extends about 350-400 km seaward, and about 400-500 km downstream. The pressure ridge on the upwind side is notably stronger than in the control case; its seaward and alongshore extent is similar to the lee side feature in this simulation, and is approaching the domain lateral boundary. Outside of the lee side wind maximum, however, surface winds over the ocean are generally weaker, which result in decreased upwelling and higher SST along the coast. The lowest temperatures differences are found at the downwind side of the tip of the cape, followed by complete separation of the upwelling front from the coastline further downstream. The upwelling front starts re-developing near the coastline, but more than 400 km downwind of the cape point, past the lee trough and wind maximum. Note that the SST changes in this simulation differ qualitatively from the flat-topography case, in location of the minimum temperatures relative to the cape, and in the inshore/offshore location of the upwelling front on the lee side of the cape.

## 7. Summary

We applied a fully two-way coupled ocean-atmosphere mesoscale system to study the atmospheric circulation around a single idealized coastal cape, simulating U.S. west coast summertime conditions featuring persistent northerly winds. This predominant northerly circulation is known to result, on one side, in wind-driven upwelling of the coastal ocean, and on the other side, in formation of localized wind maxima on the lee (downwind) side of major coastal promontories. The latter feature is also sometimes viewed as series of expansion fans and compression bulges on the downwind and windward sides of the capes, correspondingly. We found strong evidence for formation of persistent lee trough on the downwind side of an idealized cape, co-located with the wind maximum. Both wind maximum and the lee trough experienced a pronounced diurnal cycle, marked by peak in northwest flow and minimum pressure in the local evening hours, and the opposite phase of this cycle during morning hours. Weaker pressure ridge formed on the windward side of the cape, but with much lesser effect on wind regime than the lee side feature. The vertical structure of the diurnal cycle of the potential temperature and winds revealed the *downward* propagation of the temperature and wind features over the lee side coastal ocean location during the course of the day, as opposed to the traditional surface-driven development of the atmospheric boundary layer. The presence of a diurnal cycle at the surface was confirmed by the multi-year observations at buoys and coastal stations of winds and atmospheric pressure, under similar atmospheric conditions (summertime northerly winds). Satellite pictures provided further evidence of modeled cloud formation and clearing around the cape.

Temporal average of momentum analysis terms at several strategic locations around the cape confirmed the presence of strong diurnally varying local pressure gradients acting to accelerate or decelerate the boundary layer flow around the cape, depending on location and proximity to the cape point. In regions of high spatial wind variability, horizontal and vertical advection occasionally acted to balance each other. Notably reduced turbulent mixing was shown to result over the coldest SST region that had been formed on the lee side of the cape.

Analysis of relative vorticity and major terms of the vorticity equation showed strong perturbations over the coastal topography, as well as over the coastal ocean up to 200-300 km downstream and 50 km offshore. This result confirmed the hypothesis of primary influence of topography on the circulation, and the need to consider mountain flow dynamics governing the wind regime over the coastal ocean. Further sensitivity studies demonstrated critical dependence of the lee trough formation and related wind maximum upon the maximum elevation of coastal barrier. In the absence of coastal elevation, no notable lee side feature formed, while pressure ridge on the windward side of the cape was more pronounced than that on the lee side. Strongest coastal upwelling was found away from the coastal point, mainly along the straight coastline and the southern end of the cape. Progressive elevation of the coastal barrier quickly led to the formation of a well-defined lee side feature in both wind speed and sea level pressure. Additionally, higher coastal topography modified the behavior of the nearshore upwelling front: it intensified past the coastal point, but also led to greater separation from the coast

585   on the lee side of the cape. Our previous study (Perlin et al., 2011, Fig. 12b) had shown
586   that along with a band of positive wind stress curl along the coastline, a large region of
587   negative wind stress curl formed on the lee side. If positive curl favors upwelling in this
588   scenario, negative curl then leads to local downwelling, which therefore disrupts the
589   coast-following upwelling front (as resulted in no-topography simulation). The height of
590   coastal topography thus plays an important role in frequently observed separation of
591   coastal upwelling jet on the lee side of major capes, by generating local-scale negative
592   wind curl of sufficient intensity to interfere with wind-driven coastal upwelling.

593

## Acknowledgements

This research has been supported by the Office of Naval Research Grant N00014-08-1-0933. This work has been also supported in part by a grant of computer time from the DoD High Performance Computing Modernization Program at Maui High Performance Computing Center. Authors thank Roger Samelson for helpful discussions during the course of the study.

# Tables

| station | type of station | observ. record (years) | water depth/ site elevation | anemom. height | barometer height | # of days for wind analysis | # of days for Psfc analysis |
|---|---|---|---|---|---|---|---|
| 46015 | buoy | 2002-11 | 422.6 m | 5 m | sea level | 665 | 664 |
| 46027 | buoy | 2001-11 | 47.9 m | 5 m | sea level | 449 | 435 |
| 46022 | buoy | 2001-11 | 509.1 m | 5 m | sea level | 898 | 864 |
| CARO3 | coastal | 2001-11 | sea level | 8 m | 7.8 m | 800 | 655 |
| PORO3 | coastal | 2005-11 | 18.1 m ASL | 14.9 m AGL | 23.2 m ASL | 480 | 479 |
| CECC1 | coastal | 2005-11 | sea level | 8.5 m AGL | 4.8 ASL | 33* | 33* |

**Table 1.** Information on NDBC buoy and coastal stations, and the data used to estimate diurnal cycle of wind and surface pressure. AGL stands for "above ground level", ASL stands for "above mean sea level".

\* - due to the coastal location of this station, the number of days qualified for the analysis was greatly reduced.

# List of Figures

**Figure 1. a)** 14-day average 10-m horizontal wind components (vectors) and wind speed, (shading, contours, m/s). 10-m level roughly corresponds to the 4th vertical model level from the bottom. Black thick line indicates the coastline. **b)** Differences in sea surface temperature (SST) between the final time (336h) and first time record (1h) of the simulation. Contour interval is 1$^o$C; positive contours are thin black, zero and negative contours are light gray. Dashed black contours mark coastal topography for the sea level, 350m, and 750m. **c)** 14-day average sea level pressure (mb, solid contours). Dashed contours indicate ocean bathymetry for depths 2500m, 1500m, 500m, 90m and the sea level. Locations O1, O2, O3, O4, L1, L2, L3, L4 ("O" is for "ocean", "L" is for "land") are used further in text for the analysis of vertical structure. **d)** 14-day average of vertically-integrated cloud water mixing ratio (g/kg). Solid light gray line indicates location of the coastline; thick dashed dark gray lines indicate locations of vertical cross-sections for vorticity analysis in Section 5 of the manuscript.

**Figure 2.** 14-day average anomalies of 10-m wind components (vectors), wind speed magnitude (colors, m/s) and sea level pressure (contours, mb), for four distinct times of the day, 0600 LST**,** 1200 LST, 1800 LST, and 0000 LST. Anomalies are computed as departures from the corresponding daily means. Contour differences are 0.5 mb; positive contours are solid, zero and negative contours are dashed. Note that the arrows shown correspond to the vector differences and not the actual wind directions. Thus, the arrows in the same (opposite) direction of the mean flow indicate strengthening (weakening) of the wind without directional changes. For example, vectors pointing southeast (northwest) the lee side of the cape at 1800 LST (0600 LST) indicate strengthening (weakening) of a mean northwest flow in that area (compare to Fig. 1a). Similarly, the vectors pointing north (south) in the area downwind of the cape at 1200 LST (0000 LST) indicate the weakening (strengthening) of northerly component of the main wind.

**Figure 3**. (**top row**) First two EOF of the surface perturbation pressure (colors) and their local contributions into the local variance (contours, percent), for the forecast period of 36-276h; (**bottom two panels**) Amplitudes of the corresponding first and second EOFs.

**Figure 4.** 14-day average diurnal cycle of wind speed (colors, m/s) and potential temperature (contours, K) at the locations marked in Fig. 1c; **(left column)** ocean points and **(right column)** land points. Gray lines show the average diurnal cycle of the model-diagnosed PBL heights, based on Richardson number. Note that the vertical coordinate is shown as above mean sea level (ASL) height for the ocean points, and above ground level (AGL) height for the land points.

**Figure 5.** Location of buoys and coastal stations along the U.S. west near the Oregon-California border: buoys (46015, 46022, 46027) and a station CARO3 are operated by National Data Buoy Center (NDBC); PORO3, and CECC1 stations are operated by NOAA's National Ocean Service. More information on buoy and stations data is given in Table 1, and all the data are available from http://www.ndbc.noaa.gov website.

Buoy 46027 and a station CECC1 approximately represent locations downwind of the cape; buoy 46015 and a station PORO3 represent locations at the traverse of the cape; buoy 46022 and a station CARO3 represent locations upwind of the cape (Cape Mendocino for the buoy, and Cape Blanco for the station).

**Figure 6.** Average diurnal anomalies of surface winds at buoy locations and coastal stations. Time period of the observations spans May – September (upwelling-favorable months) of 2001-2011. Only days with northerly wind at all times are used for calculations, and only if the time records are complete for that calendar day. Number of days used to estimate the diurnal cycle varies for each station, and is given in Table 1.

**Figure 7.** Average diurnal anomalies of near-surface pressure at the buoy **(top panel)** and coastal stations **(bottom panel)**. Pressure anomaly is computed as the departure from the 24-h running average pressure. The gray lines denote the locations upwind of the cape(s), dashed lines correspond to the locations roughly at the traverse of the tip of the cape, and black lines correspond to the locations downwind of the cape.

**Figure 8.** Satellite images with examples of cloud clearing downstream of coastal capes and cloud build-up on the upwind side. **a)** GOES-11 satellite cloud image of southwest Oregon and northern California for 16 June 2008; b) GOES-9 image of Northern California, for 17 July 1998 (Reproduced from Fig.1 in Haack et al., 2001, by permission of the author). Note that in the first image the cloud cover extends farther offshore; the cloud coverage in the second image is limited in the offshore direction and closely follows the coastline.

**Figure 9.** 10-day average momentum balance terms for the ocean points O2, O3, O4 (**left two columns**) and land points L2, L3, L4 (**right two columns**) marked in

Fig.1c. The terms are marked as follows: **Hadv** – horizontal advection; **Vadv** – vertical advection; **Cor** – Coriolis term; **Pgeo** – geostrophic pressure gradient, imposed by the model; **Ploc** – local pressure gradient; **Vmix** – turbulent vertical mixing. For each location, left panel is for the u-components, and right panel is for the v-components. Horizontal gray line is the 10-day average model-diagnosed PBL height at the corresponding location. See text for details on pressure gradient calculations. Vertical coordinate is above sea level (ASL) height for the ocean points, and above ground level (AGL) height for the land points.

**Figure 10.** 10-day average diurnal cycle of the surface local pressure gradient terms (**Ploc** in Fig. 9) from momentum balance equations. **(left column)** u-momentum term, **(right column)** v-momentum term; **(top row)** ocean points O2, O3, O4; **(bottom row)** land points L2, L3, L4. Locations of the points are marked in Fig. 1c.

**Figure 11.** Vertical cross-sections of 13-day average vertical components of the relative vorticity (**top row**), and major terms in vorticity equation. Locations of alongshore cross-section A-B (**left column**) and cross-shore cross-section C-D (**right column**) are marked in Fig. 1d. The major vorticity equation terms are the following: (**second row**) vorticity divergence term, (**third row**) tilting and twisting term, and (**bottom row**) turbulent flux divergence term. Color scheme is for the range of values -10 to +10 of corresponding units, while the contours on the panels of vorticity terms are multiples of +/- 10.

**Figure 12.** Model results for the following cases with topography variations: **a)** flat topography (0 m); **b)** maximum coastal elevation 375 m, and **c)** maximum

coastal elevation 1500 m. **(Top row):** 14-day average 10-m horizontal wind speed (shading and thin black contours, m/s) and sea level pressure (thin gray contours). **(Bottom row):** differences in sea surface temperature (SST) between the final time (336h) and first time record (1h) of the simulation. Contour interval is $1^{o}C$; positive contours are thin black, zero and negative contours are light gray. Dashed black contours mark coastal topography for 0m, 350m, 750m, 1000m, and 1500m, when present. Corresponding fields for the control case with topography up to 750 m are shown in Fig.1 (a-b).

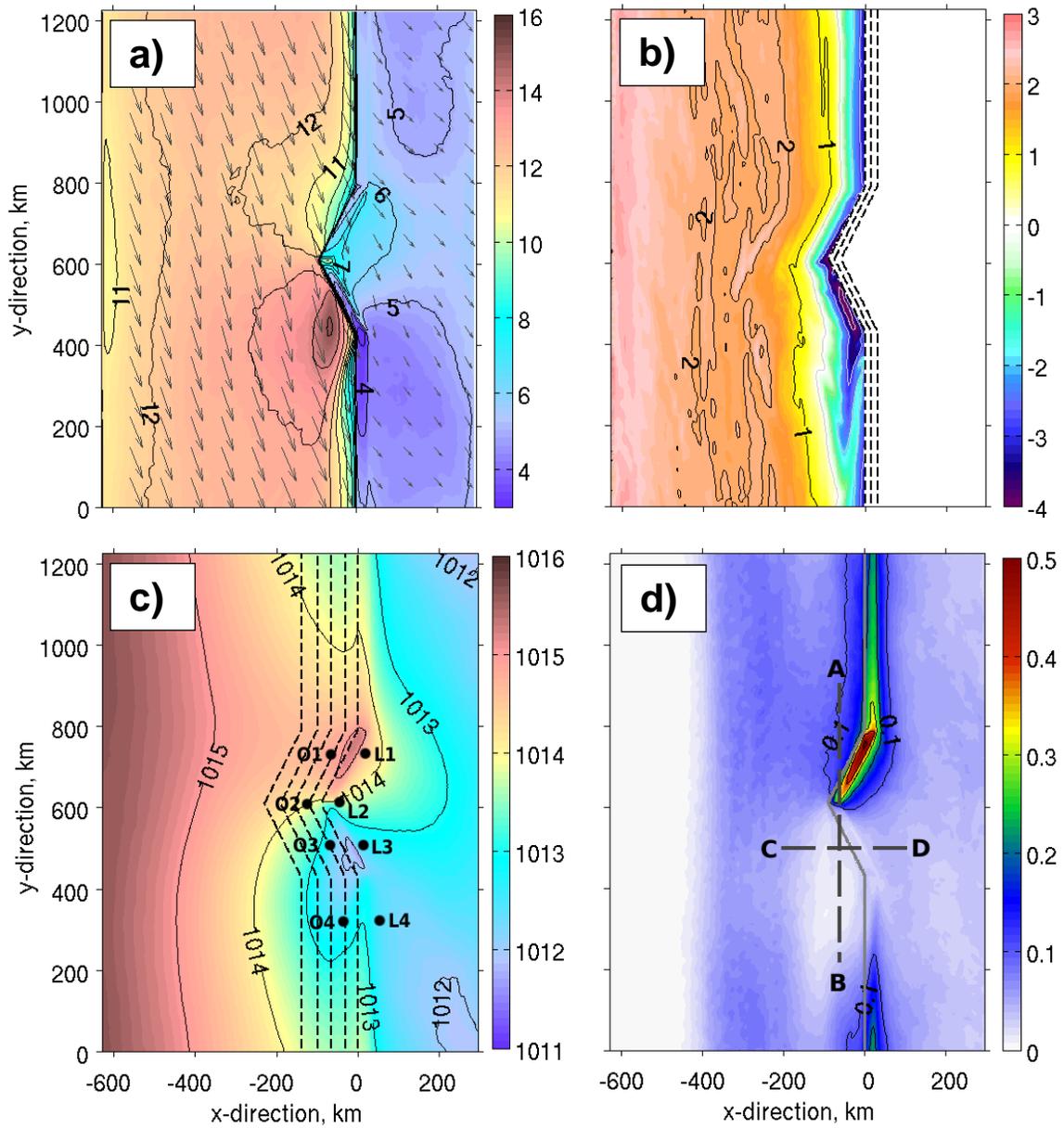

**Figure 1. a)** 14-day average 10-m horizontal wind components (vectors) and wind speed, (shading, contours, m/s). 10-m level roughly corresponds to the 4th vertical model level from the bottom. Black thick line indicates the coastline.

**b)** Differences in sea surface temperature (SST) between the final time (336h) and first time record (1h) of the simulation; SST is initially horizontally-homogeneous. Contour interval is 1°C; positive contours are thin black, zero and negative contours are light gray. Dashed black contours mark coastal topography for the sea level, 350m, and 750m.

**c)** 14-day average sea level pressure (mb, solid contours). Dashed contours indicate ocean bathymetry for depths 2500m, 1500m, 500m, 90m and the sea level. Locations O1, O2, O3, O4, L1, L2, L3, L4 ("O" is for "ocean", "L" is for "land") are used further in text for the analysis of vertical structure.

**d)** 14-day average of vertically-integrated cloud water mixing ratio (g/kg). Solid light gray line indicates location of the coastline; thick dashed dark gray lines indicate locations of vertical cross-sections for vorticity analysis in Section 5 of the manuscript.

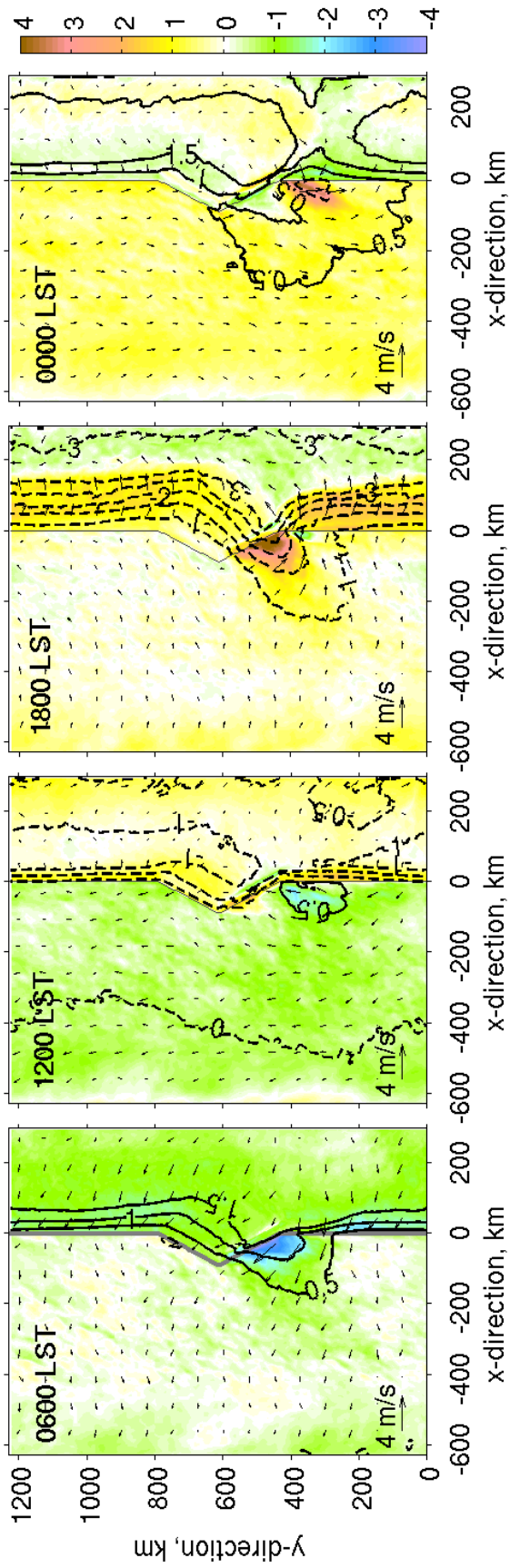

**Figure 2.** 14-day average anomalies of 10-m wind components (vectors), wind speed magnitude (colors, m/s) and sea level pressure (contours, mb), for four distinct times of the day, 0600 LST, 1200 LST, 1800 LST, and 0000 LST. Anomalies are computed as departures from the corresponding daily means. Contour differences are 0.5 mb; positive contours are solid, zero and negative contours are dashed. Note that the arrows shown correspond to the vector differences and not the actual wind directions. Thus, the arrows in the same (opposite) direction of the mean flow indicate strengthening (weakening) of the wind without directional changes. For example, vectors pointing southeast (northwest) the lee side of the cape at 1800 LST (0600 LST) indicate strengthening (weakening) of a mean northwest flow in that area (compare to Fig. 1a). Similarly, the vectors pointing north (south) in the area downwind of the cape at 1200 LST (0000 LST) indicate the weakening (strengthening) of northerly component of the main wind.

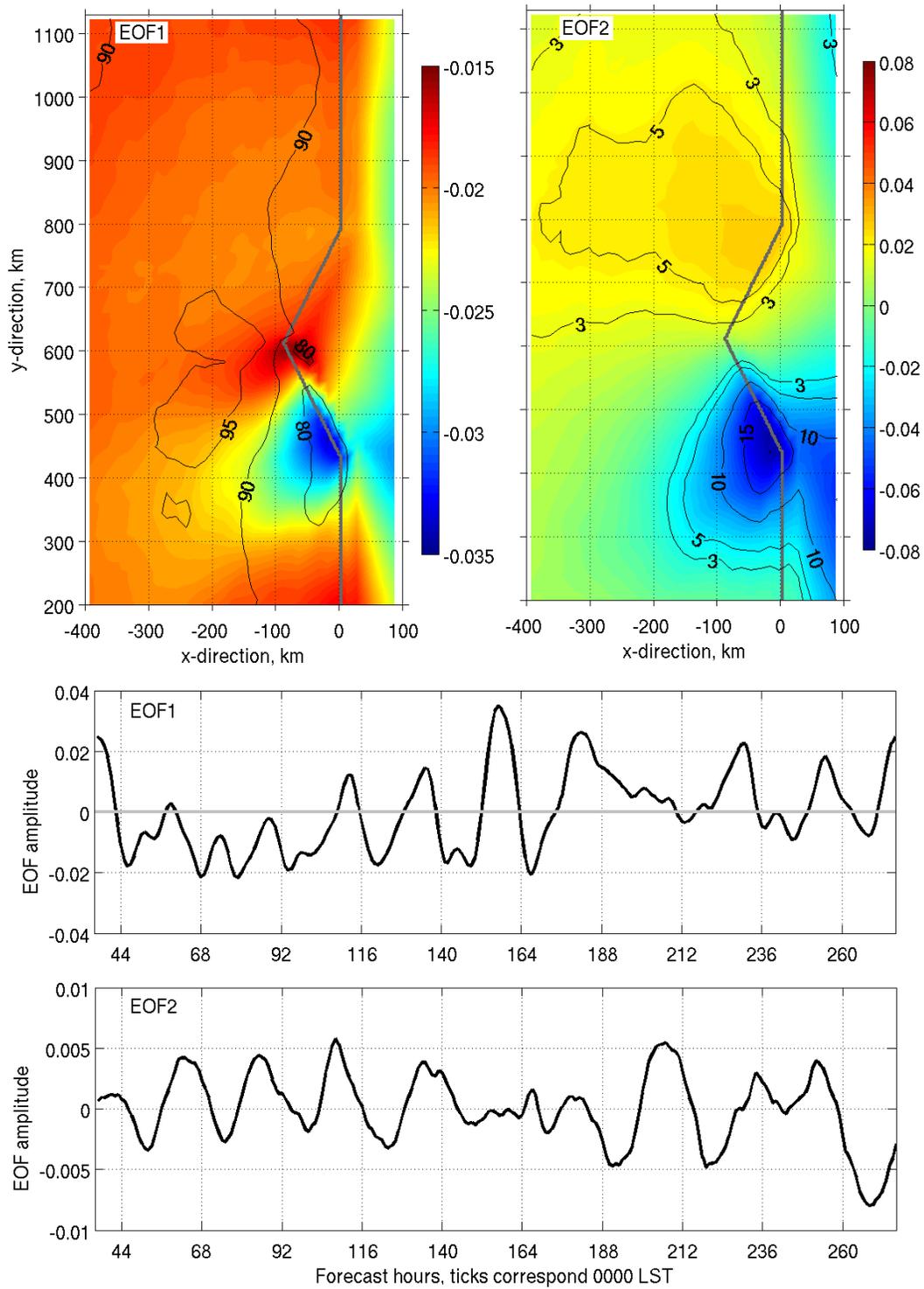

**Figure 3**. (**top row**) First two EOF of the surface perturbation pressure (colors) and their local contributions into the local variance (contours, percent), for the forecast period of 36-276h; (**bottom two panels**) Amplitudes of the corresponding first and second EOFs.

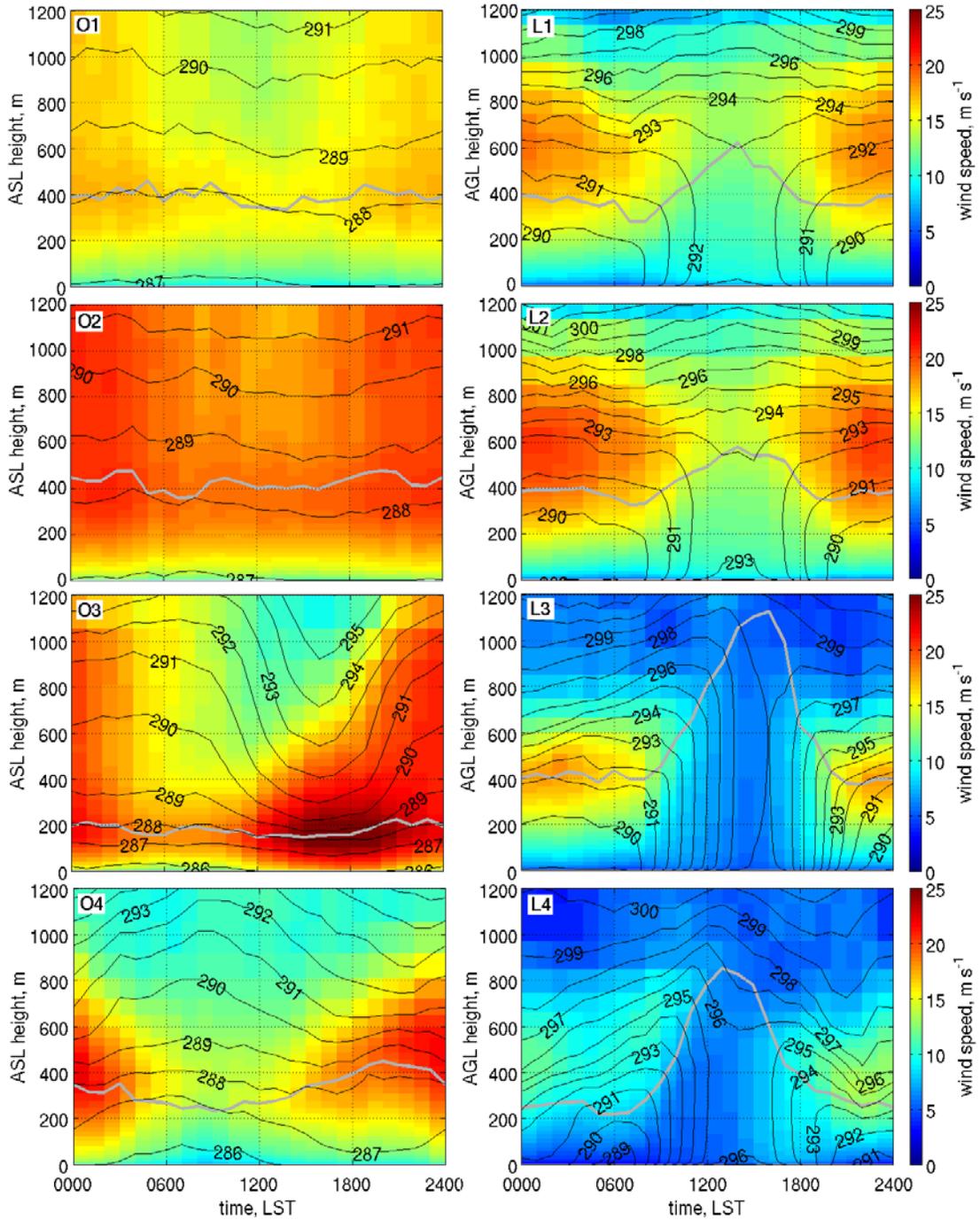

**Figure 4.** 14-day average diurnal cycle of wind speed (colors, m/s) and potential temperature (contours, K) at the locations marked in Fig. 1c; **(left column)** ocean points and **(right column)** land points. Gray lines show the average diurnal cycle of the model-diagnosed PBL heights, based on Richardson number. Note that the vertical coordinate is shown as above mean sea level (ASL) height for the ocean points, and above ground level (AGL) height for the land points.

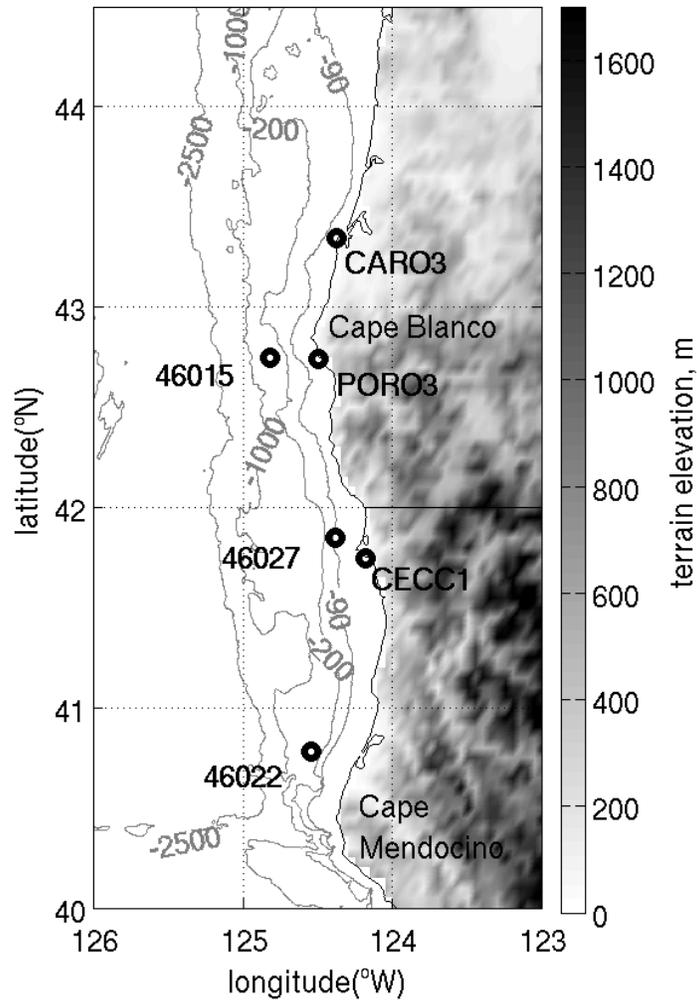

**Figure 5.** Location of buoys and coastal stations along the U.S. west near the Oregon-California border: buoys (46015, 46022, 46027) and a station CARO3 are operated by National Data Buoy Center (NDBC); PORO3, and CECC1 stations are operated by NOAA's National Ocean Service. More information on buoy and stations data is given in Table 1, and all the data are available from http://www.ndbc.noaa.gov website.

Buoy 46027 and a station CECC1 approximately represent locations downwind of the cape; buoy 46015 and a station PORO3 represent locations at the traverse of the cape; buoy 46022 and a station CARO3 represent locations upwind of the cape (Cape Mendocino for the buoy, and Cape Blanco for the station).

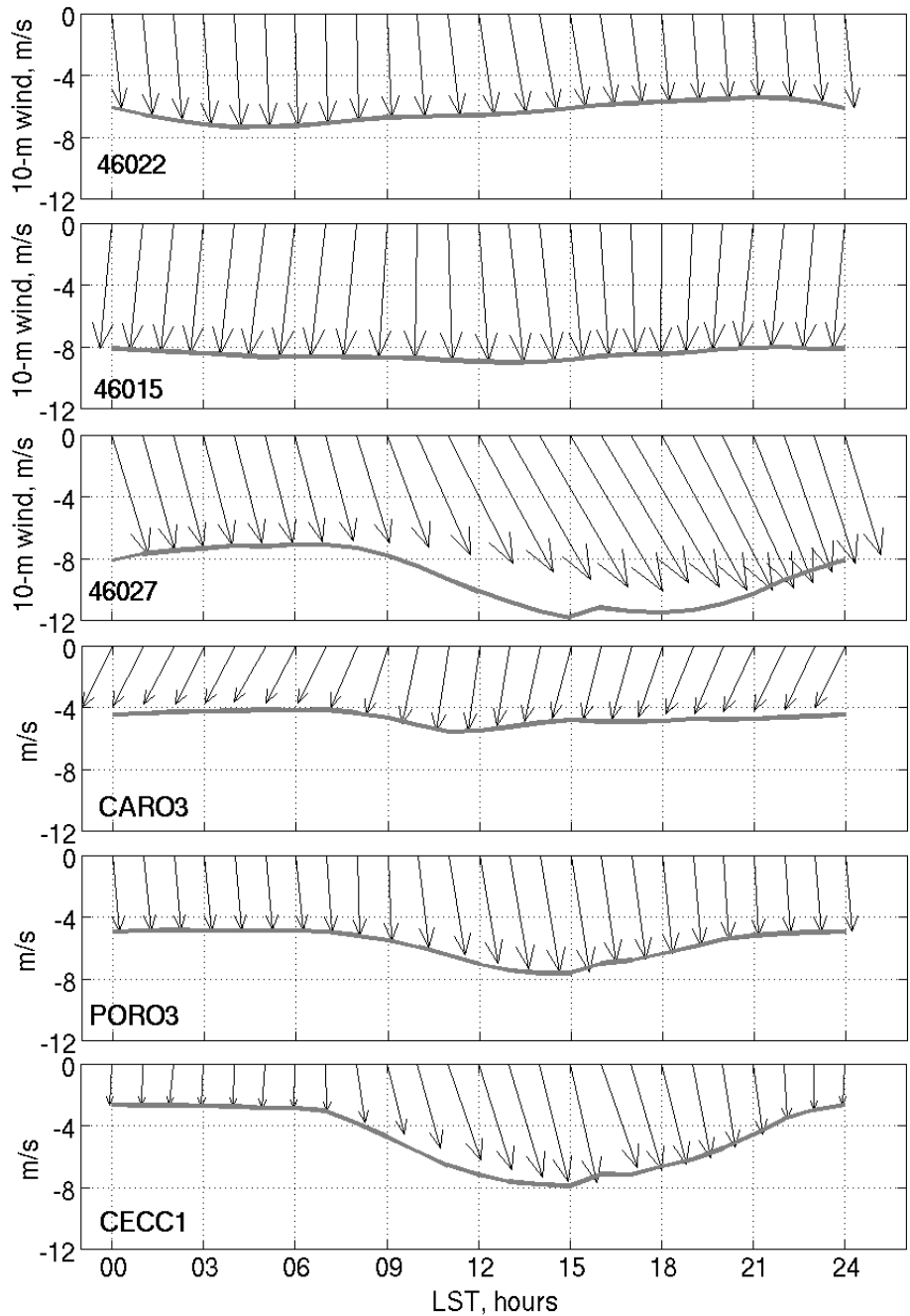

**Figure 6.** Average diurnal cycle of surface winds at buoy locations and coastal stations; gray lines outline wind speed (with a negative sign). Time period of the observations spans May – September (upwelling-favorable months) of 2001-2011. Only days with northerly wind at all times are used for calculations, and only if the time records are complete for that calendar day. Number of days used to estimate the diurnal cycle varies for each station, and is given in Table 1.

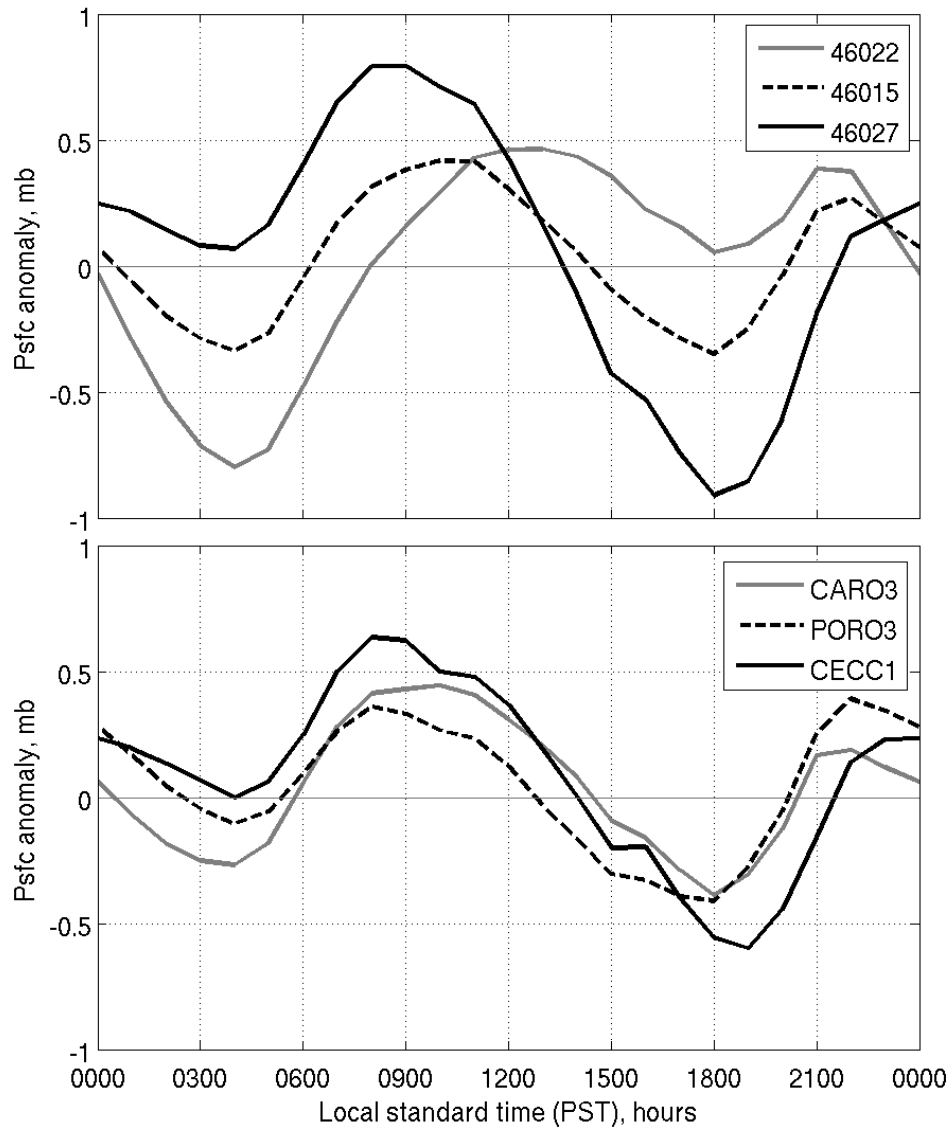

**Figure 7.** Average diurnal anomalies of near-surface pressure at the buoy **(top panel)** and coastal stations **(bottom panel)**. Pressure anomaly is computed as the departure from the 24-h running average pressure. The gray lines denote the locations upwind of the cape(s), dashed lines correspond to the locations roughly at the traverse of the tip of the cape, and black lines correspond to the locations downwind of the cape.

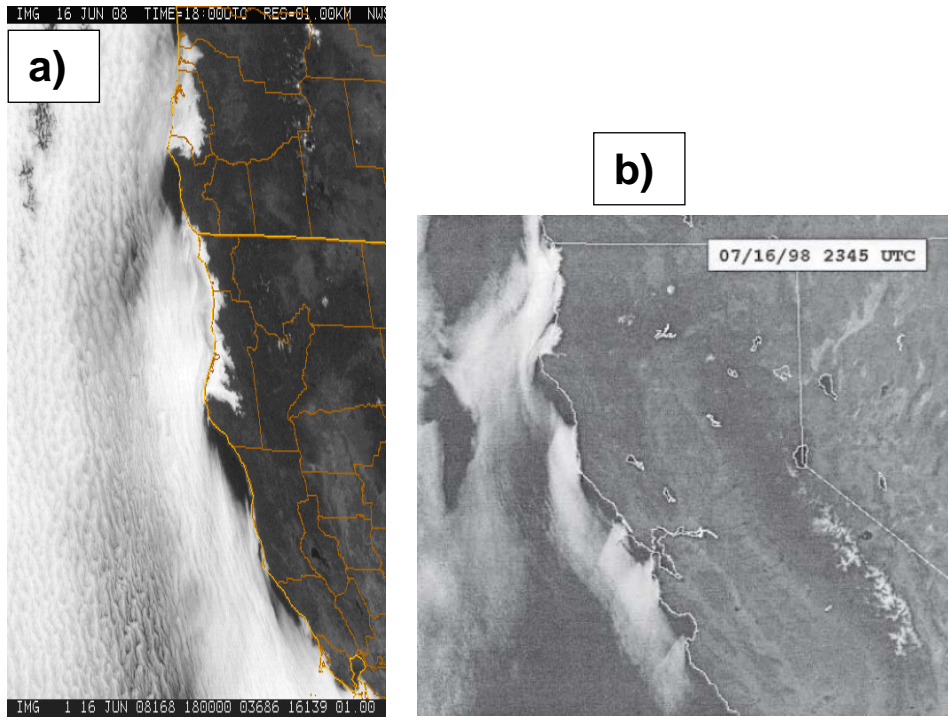

**Figure 8.** Satellite images with examples of cloud clearing downstream of coastal capes and cloud build-up on the upwind side. **a)** GOES-11 satellite cloud image of southwest Oregon and northern California for 16 June 2008; b) GOES-9 image of Northern California, for 17 July 1998 (Reproduced from Fig.1 in Haack et al., 2001, by permission of the author). Note that in the first image the cloud cover extends farther offshore; the cloud coverage in the second image is limited in the offshore direction and closely follows the coastline.

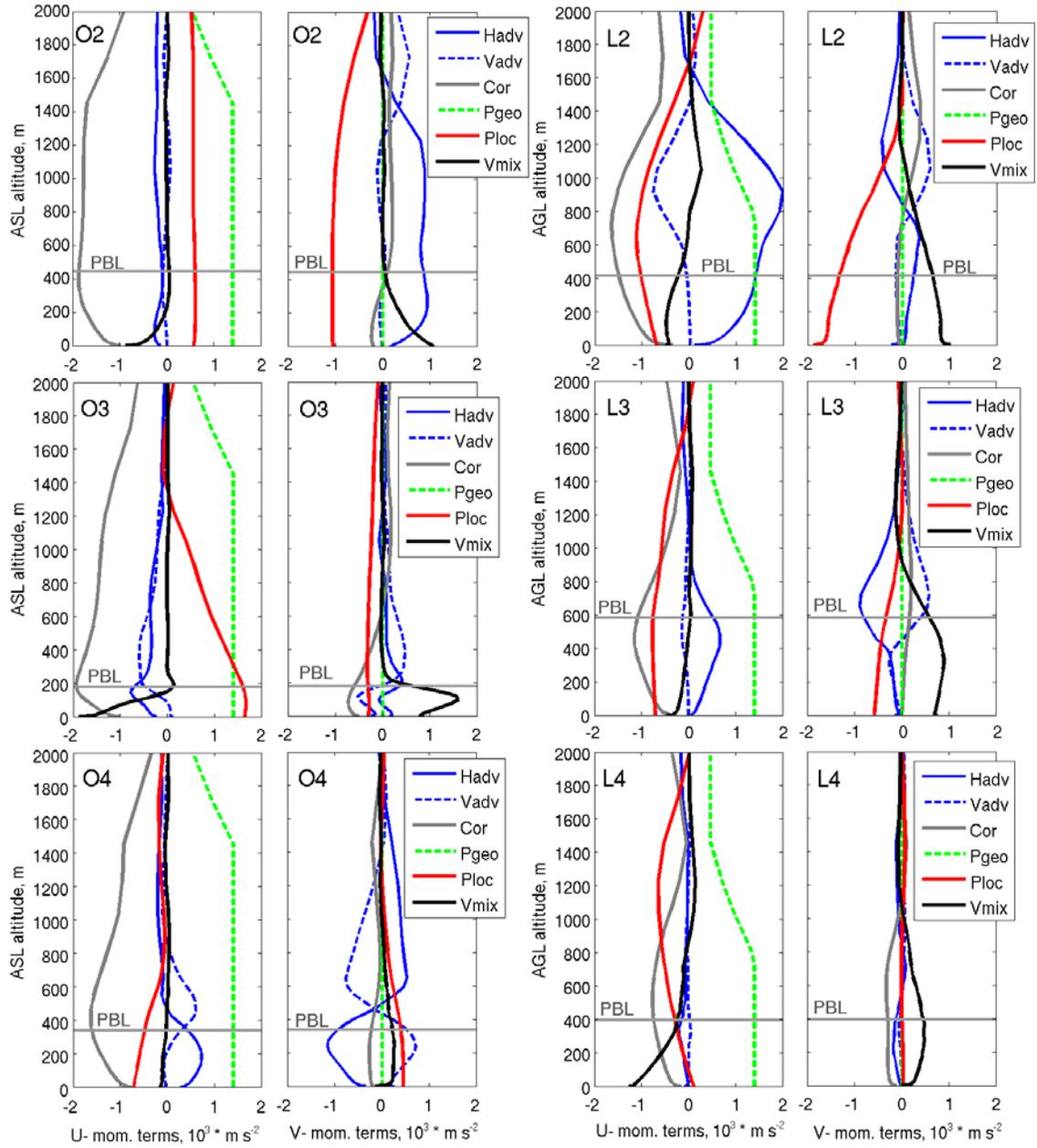

**Figure 9.** 10-day average momentum balance terms for the ocean points O2, O3, O4 (**left two columns**) and land points L2, L3, L4 (**right two columns**) marked in Fig. 1c. The terms are marked as follows: **Hadv** – horizontal advection; **Vadv** – vertical advection; **Cor** – Coriolis term; **Pgeo** – geostrophic pressure gradient, imposed by the model; **Ploc** – local pressure gradient; **Vmix** – turbulent vertical mixing. For each location, left panel is for the u-components, and right panel is for the v-components. Horizontal gray line is the 10-day average model-diagnosed PBL height at the corresponding location. See text for details on pressure gradient calculations. Vertical coordinate is above sea level (ASL) height for the ocean points, and above ground level (AGL) height for the land points.

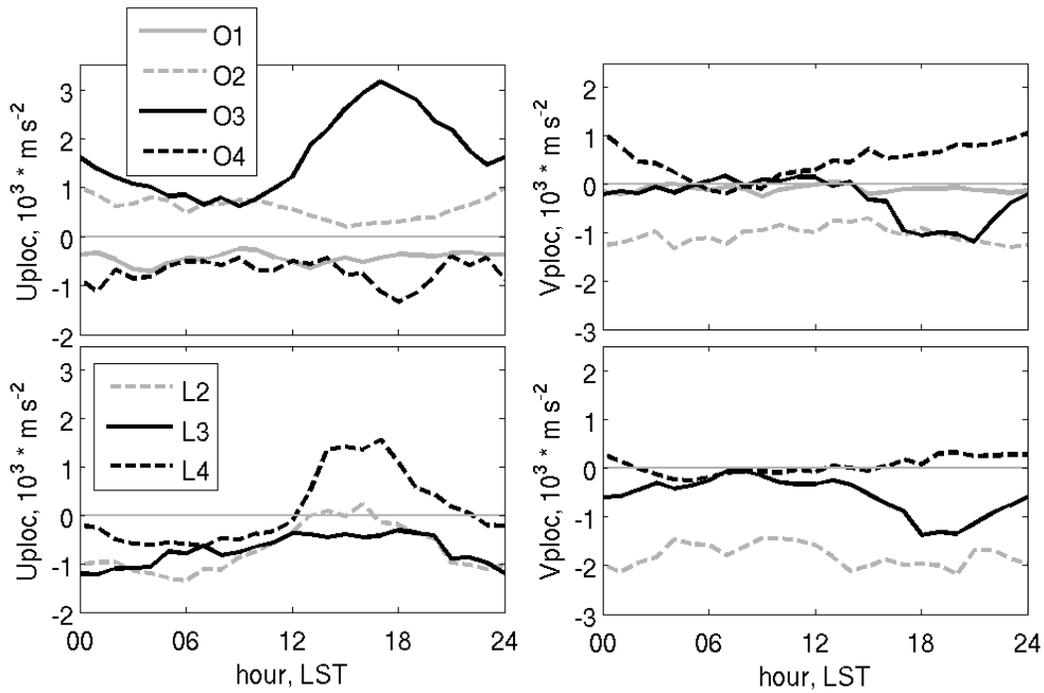

**Figure 10.** 10-day average diurnal cycle of the surface local pressure gradient terms (**Ploc** in Fig. 9) from momentum balance equations. **(left column)** u-momentum term, **(right column)** v-momentum term; **(top row)** ocean points O2, O3, O4; **(bottom row)** land points L2, L3, L4. Locations of the points are marked in Fig.1c.

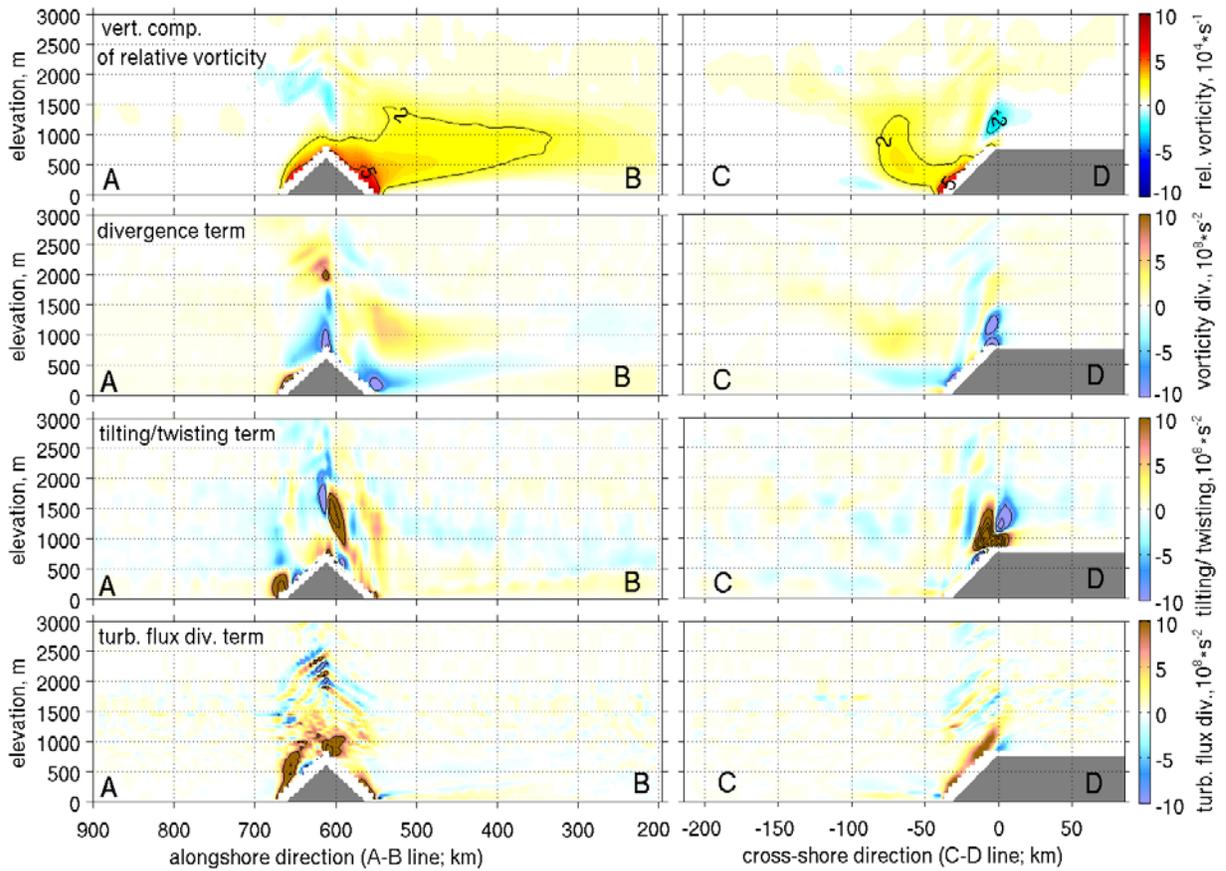

**Figure 11.** Vertical cross-sections of 13-day average vertical components of the relative vorticity (**top row**), and major terms in vorticity equation. Locations of alongshore cross-section A-B (**left column**) and cross-shore cross-section C-D (**right column**) are marked in Fig. 1d. The major vorticity equation terms are the following: (**second row**) vorticity divergence term, (**third row**) tilting and twisting term, and (**bottom row**) turbulent flux divergence term. Color scheme is for the range of values -10 to +10 of corresponding units, while the contours on the panels of vorticity terms are multiples of +/- 10.

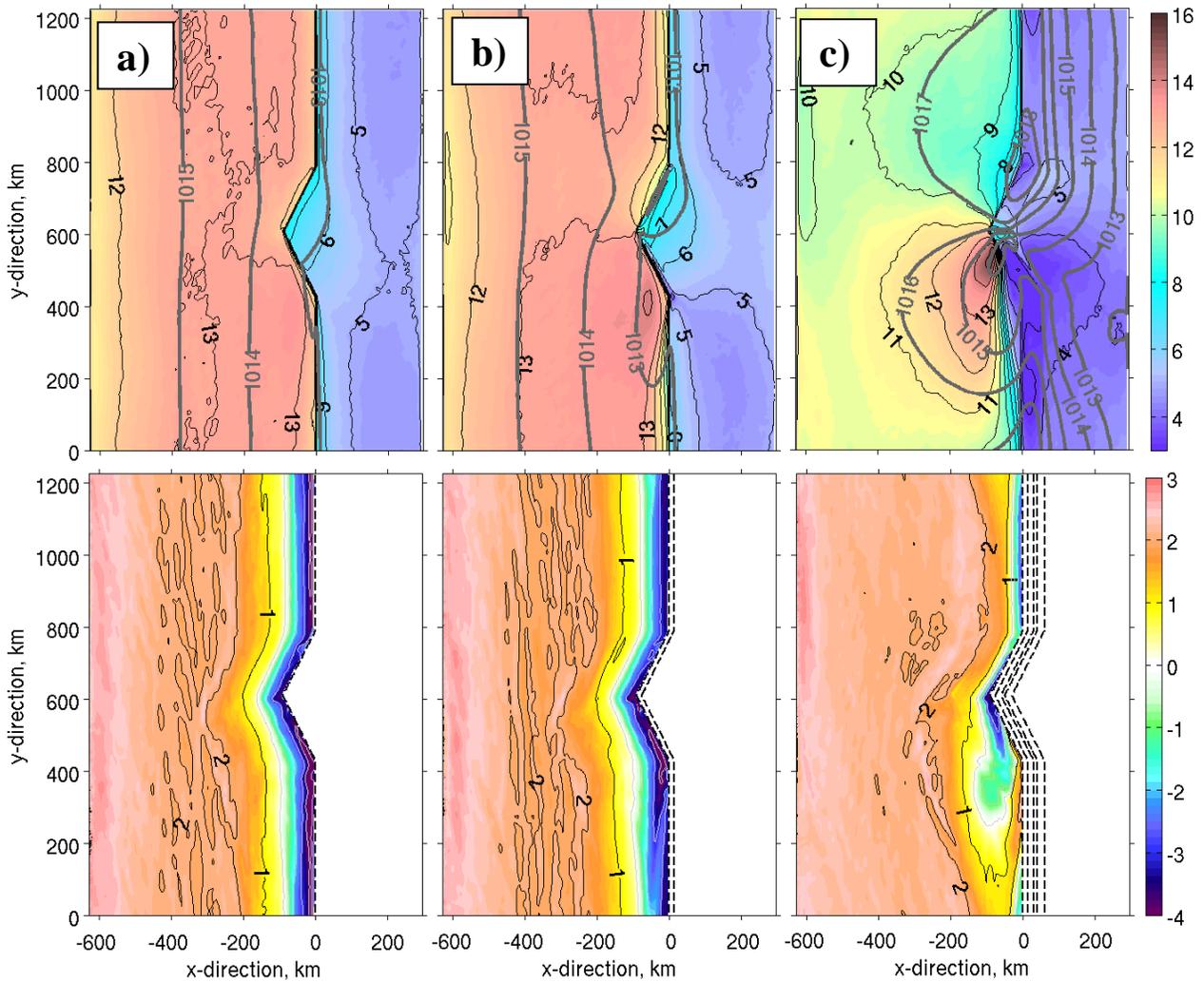

**Figure 12.** Model results for the following cases with topography variations: **a)** flat topography (0 m); **b)** maximum coastal elevation 375m, and **c)** maximum coastal elevation 1500m. **(Top row):** 14-day average 10-m horizontal wind speed (shading and thin black contours, m/s) and sea level pressure (thin gray contours). **(Bottom row):** differences in sea surface temperature (SST) between the final time (336h) and first time record (1h) of the simulation. Contour interval is 1°C; positive contours are thin black, zero and negative contours are light gray. Dashed black contours mark coastal topography for 0m, 350m, 750m, 1000m, and 1500m, when present. Corresponding fields for the control case with topography up to 750 m are shown in Fig.1 (a-b).

53